\documentclass{cernrep} 
\usepackage{texnames}
\usepackage[T1]{fontenc}
\usepackage{hyperref} 
 \def\bc{\begin{center}}          \def\ec{\end{center}}
\usepackage{subfig}
\usepackage{graphicx}
\usepackage{rotating}
\usepackage{float}

\usepackage{units}

\begin{document}

\begin{center}

\includegraphics[width=0.4\textwidth]{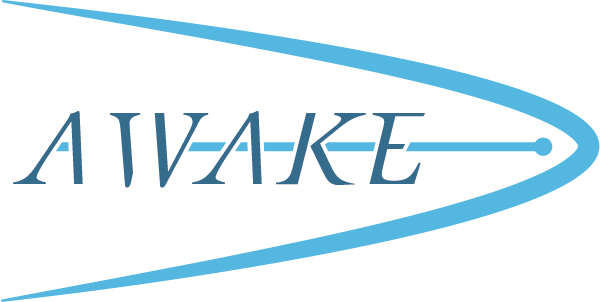}

{\LARGE Input to the European Strategy for Particle Physics Update
\vskip 0.3 cm 
on behalf of the AWAKE Collaboration}

\vspace{1cm}
{Contact persons: E.~Gschwendtner\footnote{edda.gschwendtner@cern.ch}, P.~Muggli\footnote{muggli@mpp.mpg.de}, M.~Turner\footnote{marlene.turner@cern.ch}}
\end{center}
\vspace{0.5cm}
\begin{center}
{March 31, 2025}
\end{center}

\vspace{0.5cm}

\textit{Abstract} -- The Advanced Wakefield Experiment, AWAKE, is a well-established international collaboration and aims to develop the proton-driven plasma wakefield acceleration of electron bunches to energies and qualities suitable for first particle physics applications, such as strong-field QED and fixed target experiments (\unit[$\sim$50–200]{GeV}).
Numerical simulations show that these energies can be reached with an average accelerating gradient of \unit[$\sim1$]{GeV/m} in a single proton-driven plasma wakefield stage. %
This is enabled by the high energy per particle and per bunch of the CERN SPS (\unit[$\sim$19]{kJ}, \unit[400]{GeV}) and LHC (\unit[$\sim$120]{kJ}, \unit[7]{TeV}) proton bunches. %
Bunches produced by synchrotrons are long, and AWAKE takes advantage of the self-modulation process to drive wakefields with GV/m amplitude. 

By the end of 2025, all physics concepts related to self-modulation will have been experimentally established as part of the AWAKE ongoing program that started in 2016. %
Key achievements include: direct observation of self-modulation, stabilization and control by two seeding methods, acceleration of externally injected electrons from \unit[19]{MeV} to more than \unit[2]{GeV}, and sustained high wakefield amplitudes beyond self-modulation saturation using a plasma density step. %

In addition to a brief summary of achievements reached so far, this document  outlines the AWAKE roadmap as a demonstrator facility for producing beams with quality sufficient for first applications. %
The plan includes:  
\begin{itemize}
    \item Accelerating a quality-controlled electron bunch to multi-GeV energies in a \unit[10]{m} plasma by 2031; %
    \item Demonstrating scalability to even higher energies by LS4. %
\end{itemize}
 
Synergies of the R\&D performed in AWAKE that are relevant for advancing plasma wakefield acceleration in general are highlighted. 

We argue that AWAKE and similar advanced accelerator R\&D be strongly supported by the European Strategy for Particle Physics Update.

\newpage

\section{The AWAKE Collaborating Institutes}
\begin{itemize}
    \item CERN, Geneva, Switzerland
    \item Cockcroft Institute, Daresbury, UK
    \item GoLP/Instituto de Plasmas e Fusão Nuclear, Instituto Superior Técnico, Universidade de Lisboa, Lisbon, Portugal
    \item Heinrich-Heine-University of Düsseldorf, Düsseldorf, Germany
    \item Hungarian Research Network, HUN-REN, Budapest, Hungary
    \item Imperial College, London, UK
    \item ISCTE - Instituto Universitéario de Lisboa, Portugal
    \item Max Planck Institute for Physics, Garching/Munich, Germany
    \item Max Planck Institute for Plasma Physics, Greifswald, Germany
    \item Philipps-Universität Marburg, Marburg, Germany
    \item Swiss Plasma Center, EPFL, Lausanne, Switzerland 
    \item University College London, London, UK
    \item University of Lancaster, Lancaster, UK
    \item University of Liverpool, Liverpool, UK
    \item University of Manchester, Manchester, UK
    \item University of Oxford, Oxford, UK
    \item University of Wisconsin, Madison, US
    \item UNIST, Ulsan, Republic of Korea
    \item Uppsala University, Uppsala, Sweden
\end{itemize}

\section{Introduction}
\subsection{Motivation}
The key motivation for plasma wakefield acceleration is high gradient ($>$GeV/m) particle acceleration enabling more compact future linear accelerators. %
The concept of using relativistic particle bunches to drive high-amplitude wakefields in plasma was proposed in 1985~\cite{chen}, and experimentally demonstrated soon after~\cite{rosenzweig}. %
The highest energy gain to date -- \unit[42]{GeV} over \unit[85]{cm} -- was achieved at SLAC using a single \unit[42]{GeV} electron bunch~\cite{blumenfeld}. %
Subsequent experiments focused on high energy transfer efficiency~\cite{litos}. %
More recently, emittance preservation, albeit with limited energy gain~\cite{lindstrom} was demonstrated at DESY. %

Reaching very high gradients (\unit[$\sim$50]{GeV}), as in the SLAC experiment, requires short and tightly focused (on the order of \unit[10]{$\mu$m}), dense (beam density $n_b$ on the order of the plasma electron density $n_{pe}$) driver bunches together with a high plasma density (on the order of \unit[$\sim 10^{17}$]{cm$^{-3}$}). %
In such a configuration, the energy gain is typically limited by the incoming drive particle energy, which limits the distance over which the wakefields can be excited~\footnote{For example, based on the conservation of energy, a \unit[50]{GeV} driver can excite wakefield with an amplitude of \unit[50]{GeV/m} over approximately \unit[1]{m}}. %

This limit motivates the use of energetic proton drive bunches, such as the ones produces at the CERN Synchrotrons, the SPS (\unit[400]{GeV}) and LHC (\unit[7]{TeV}) and offers the benefit of delivering sufficient energy to accelerate a witness bunch to the levels necessary for a Higgs Factory and beyond, in a single plasma stage. %
This capability eliminates the need for staging, thereby significantly simplifying the design and operation of the accelerator complex. %

However, bunches produced by the current proton machines are long (\unit[$\sigma_z>5$]{cm}), and can excite only low-amplitude (\unit[$\sim10$]{MV/m}) wakefields unless properly prepared, for example by self-modulation in plasma. %

Using proton bunches as wakefield drivers is the idea underlying the AWAKE experiment at CERN. %
In AWAKE, the long and tightly focused bunches provided by the Super Proton Synchrotron (SPS) undergo self-modulation~\cite{kumar} in a low-density plasma. %
Started by initial low-amplitude wakefields (seed wakefields), the bunch forms into a train of microbunches, each shorter than, and spaced by, the plasma wavelength. %
Driven by the forming microbunch train, the wakefield amplitude grows along the plasma from the seed wakefields to saturation (\unit[30-50]{\%} of the wavebreaking field)~\cite{lotovEwb}.
In a uniform plasma, the wakefield amplitude decreases after saturation due to continuous evolution of the bunch train. %
However, numerical simulation results show that introducing a small (few percent) plasma density step at the correct location stabilizes the wakefield amplitude over long distances~\cite{lotovstep}. %
Simulations also indicate that an electron bunch injected on-axis can preserve its emittance and, with proper wakefield loading, maintain a narrow energy spread (percent-level) while accelerating~\cite{veronica}.

Based on these considerations the AWAKE collaboration follows a clear roadmap towards proposing first particle physics applications using the proton-driven plasma wakefield acceleration of electrons to high energies. 
\subsection{Timeline and Program}
    \begin{figure}[phtb]
    \centering
    \includegraphics[scale=0.7]{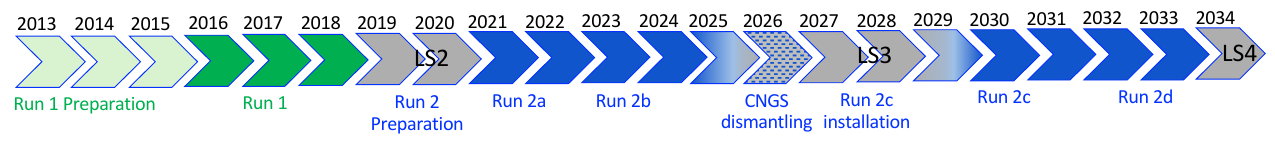}
    \caption{Timeline of the AWAKE experimental program.%
    } 
    \label{fig:AWAKE-timeline} 
    \end{figure}

AWAKE is an international collaboration, consisting of 19 institutes world-wide and was approved in 2013. %
The timeline of the AWAKE experimental program is shown in Fig.~\ref{fig:AWAKE-timeline}, starting with first experiments in 2016. %
This program has been highly successful, with firstly demonstrating the proof-of-concept of seeded self-modulation of the proton bunch and the acceleration of externally injected electrons from 19\,MeV to 2\,GeV (Run 1) and secondly with the development of the self-modulator plasma completed by the end of 2025 (Run 2a/b), with the start of CERN's Long Shutdown 3 (LS3). %
For the post-LS3 period (Run 2c/d) from 2029 until 2033, AWAKE plans to focus work on the accelerator and the production of multi-GeV bunches with quality. %
For that, a \unit[150]{MeV} electron linac, using advanced S- and X-band structures, is being developed for electron injection into the accelerator plasma. %

During LS3, the CERN Neutrino Gran Sasso, CNGS, target area will be cleaned out in order to expand the AWAKE area, that is installed just upstream of it. %
Moreover, the facility will be upgraded to accommodate the new, 150\,MeV linac and the second \unit[10]{m} plasma source used for acceleration, with the flexibility to support scalable plasma sources up to \unit[50]{m}. %

To enable further scaling (hundreds of meters to kilometers), AWAKE is developing two scalable plasma source technologies: discharge plasma, and helicon plasma. %
One of these is expected to be ready for the second half of the post-LS3 program. %

The AWAKE plan has received strong support from an international review committee mandated by CERN,~\footnote{\href{https://edms.cern.ch/document/3038236/0.1}{https://edms.cern.ch/document/3038236/0.1}} and was approved by the CERN Council in 2024. %

After successful completion of this program, AWAKE will be in a position to propose particle physics applications with accelerated energy bunch energies suitable for strong-field QED and fixed-target experiments (\unit[$\sim$50–200]{GeV})~\cite{PBC, symmetry}.
Several collider studies such as electron-proton~\cite{symmetry} and electron-positron colliders~\cite{alive} at the TeV levels are under way, which are based on the successful demonstration of the AWAKE technology. 

\section{AWAKE Results}

\begin{figure}[phtb]
    \centering
    \includegraphics[scale=0.8]{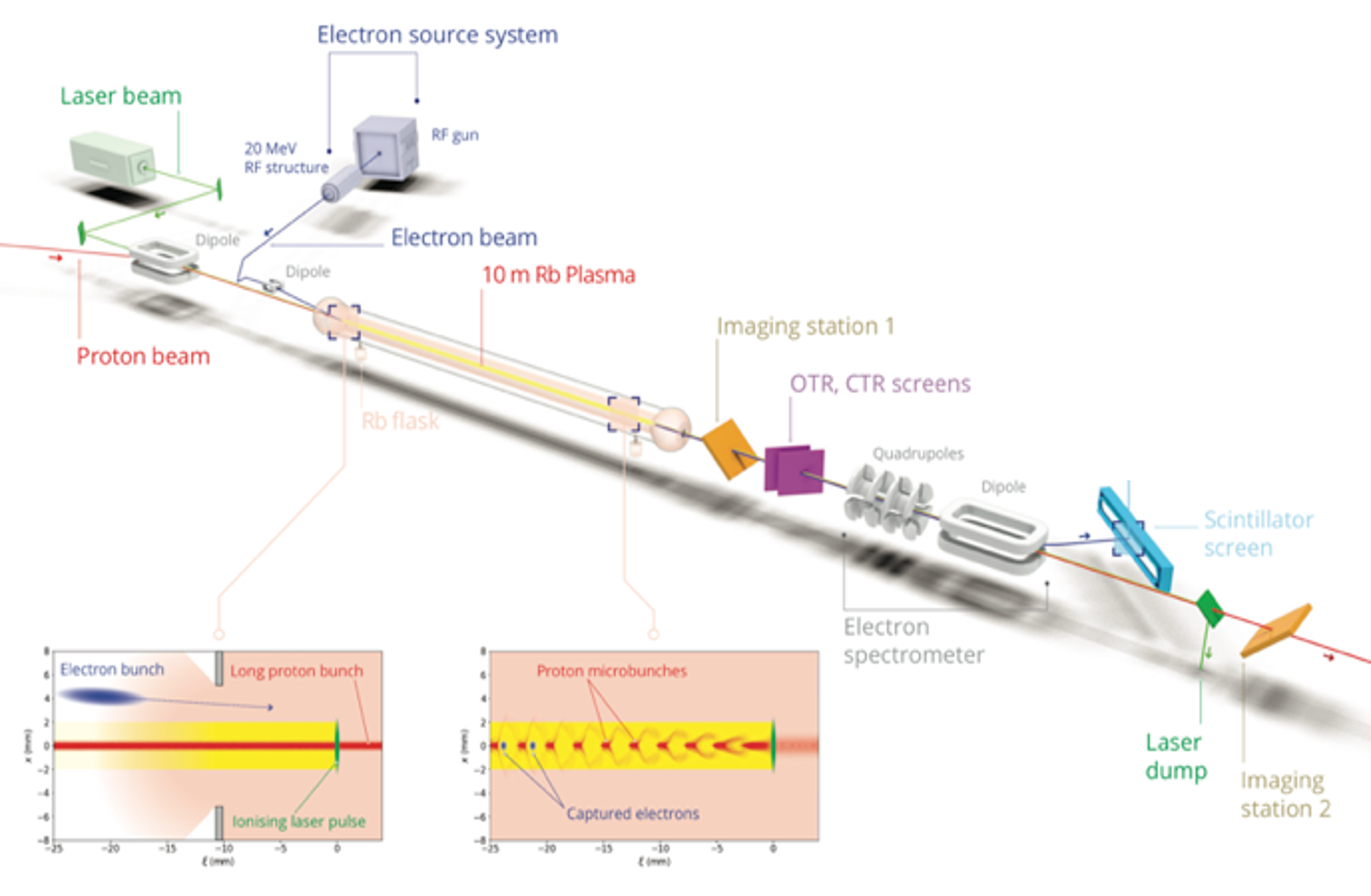}
    \caption{Layout of the experimental setup of AWAKE from 2016 until end 2025, showing in particular the single vapor/plasma source (or discharge source (DPS)), and the electron  source for seeding and acceleration experiments. %
    Insets show the side-injection scheme, and the self-modulated bunch and accelerated electrons in the plasma. %
    } 
    \label{fig:LayoutRun1} 
    \end{figure}

\begin{figure}[phtb]
    \centering
    \includegraphics[scale=0.6]{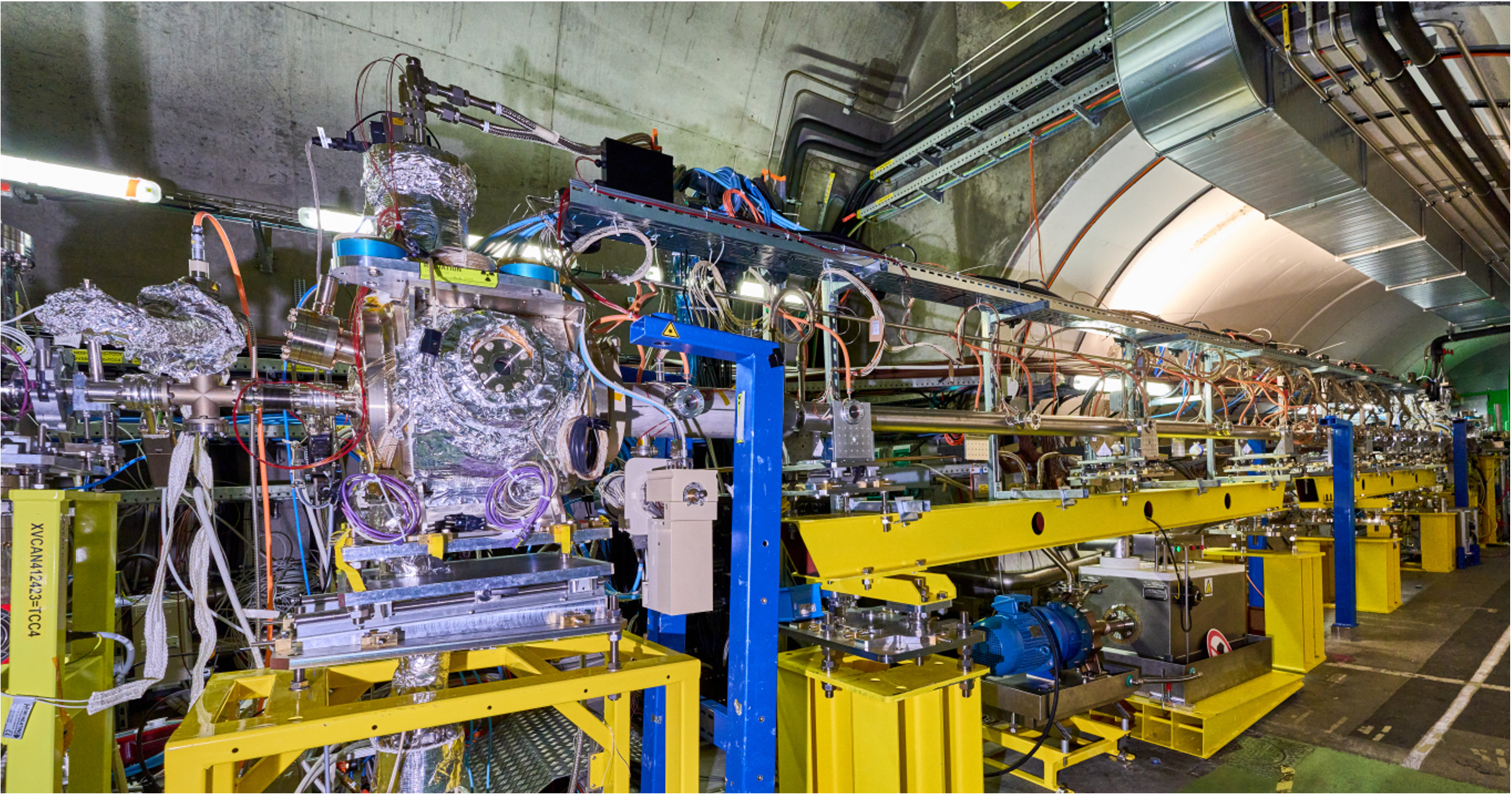}
    \caption{Latest 10\,m-long vapor/plasma source installed in the AWAKE facility. %
    It includes the option to impose a step in temperature and thus in vapor and plasma densities, and to vary the plasma length by inserting foils that block the ionizing laser pulse. %
    } 
    \label{fig:VaporSourceInstalled} 
    \end{figure}

 We summarize here the results obtained from the start of the experimental program in 2016 until today, which are the basis for the future of AWAKE. %
 These were obtained with the experimental setup shown in Fig.~\ref{fig:LayoutRun1}. %
    Figure~\ref{fig:VaporSourceInstalled} shows the latest vapor/plasma rubidium source installed in the AWAKE facility. %
    It includes the option to impose a step in plasma density and to vary the plasma length. %

\subsection{Self-Modulation}
While self-modulation, SM, of a long laser pulse was the method used to drive plasma wakefields before the advent of ultra-short (fs) laser pulses, it had never been directly observed, nor had it been observed with a particle bunch. %
In AWAKE we showed for the first time that SM produces the expected microbunch train of the proton bunch, i.e., with the period of the wakefields, over a range of ten in plasma density~\cite{karl,marlene}. %
Measurements also show that the amplitude of transverse wakefields increases from an initially low amplitude to reach hundreds of MV/m, with defocused particles forming a halo on time-integrated images of the transverse bunch distribution. %
Observation of SM is evidence of the driving of transverse wakefields. %

SM develops from the self-interaction between the proton bunch and transverse wakefields. %
Therefore, SM can develop as an instability, leading to different outcomes from event to event when developing from noise~\cite{lotovnoise} or imperfections in the bunch density profile~\cite{fabian}. %
However, controlled initial wakefields can be imposed as a seed to make the process reproducible~\cite{fabian,livio}. %
The first method we used to seed the SM process involves a short laser pulse, propagating colinearly with the proton bunch and within the bunch, that ionizes the rubidium (Rb) vapor. %
The sudden onset of the beam-plasma interaction at this relativistic ionization front (RIF) provides seed wakefields with an amplitude proportional to the bunch density at the RIF location. %
When their amplitude is sufficient to overcome other ''noise'' wakefields, these seeded wakefields grow reaching a fixed amplitude and phase; they become reproducible. %
The ability to seed wakefields and the transition between growth as an instability to a seeded process were demonstrated using the RIF~\cite{fabian}. %

The ability to seed SM was also demonstrated by placing a short electron bunch behind the RIF, but much ahead of the proton bunch (thus not seeded by the RIF)~\cite{livio}. %
This seeding method has advantages when compared to the RIF method. %
It allows for independent control of the seed level, which depends only on the electron bunch parameters, and on the growth rate of SM, which in turn depends on the proton bunch parameters (for a fixed plasma density). %
Additionally, this method enables the self-modulation of the entire proton bunch, not just a fraction of it. %

Seeding is essential to deterministically injecting the electron bunch at a precise location where the wakefields have maximum amplitude, and are both accelerating and focusing~\cite{veronica}. %
This is accomplished by choosing the delay between the seed (RIF or electron bunch) and the injected bunch, which can then be easily optimized experimentally. %

\subsection{Electron Acceleration within the Self-Modulator}

We externally injected test electrons into the wakefields and measured acceleration from the initial 19\,MeV energy to about 2\,GeV~\cite{nature,royal}, at a plasma density near to the predicted optimum (7$\times$10$^{14}$\,cm$^{-3}$). %
This is evidence for the excitation of longitudinal wakefields. %
For these experiments, injection occurred from the side and at locations 3-7\,m after the plasma entrance to prevent the electrons from being lost in the regions where wakefields develop. %
Indeed, in the density ramp at the plasma entrance, the proton bunch drives strong transverse fields that cause the electrons to defocus~\cite{pablothesis} and be lost. %
Additionally, until saturation of the wakefields along the plasma, the wakefields have a phase velocity slower than the velocity of the bunch, and again, electrons would be defocused because of relative dephasing~\cite{pukhov,schroeder}. %
Evidence of this slow phase velocity was observed when applying linear density gradients along the plasma~\cite{falk,pablo}. %
Density gradients modify the phase velocity of the wakefields, thereby affecting their development. %

In 2024, we upgraded the vapor source to allow varying the length of the plasma. %
This was achieved by adding insertable foils, that block the RIF at various locations along the source (i.e., 0.5.1.5, ..,9.5\,m). %
This allows for measurements of energy gain (and many other parameters) over the last meters of the plasma, making it possible to calculate the accelerating gradient averaged over one meter. %
These measurements have successfully demonstrated that, with plasma of constant density, the accelerating gradient decreases over the last 4-5\,m of plasma, as expected. %
They also show that, by introducing a density step, the gradient can be maintained at a higher value over that distance. %
The quality of these measurements is currently limited by the side-injection process, required with the self-modulator. %
However, these measurements will be repeated and improved once the two-plasma system will become available. %

\subsection{Other Results}
The results outlined above form the foundation of the self-modulator, and will enable acceleration to high energies in a second plasma. %
Additionally, we have acquired other important results that enhance our understanding and control of the beam plasma system. %
These results were published in high-level journals, including:
\begin{itemize}
    \item \textit{Experimental Observation of Motion of Ions in a Resonantly Driven Plasma Wakefield Accelerator}, accepted in \textbf{Phys. Rev. Lett.} (Feb. 2025)~\cite{marleneionmotion}
    \item \textit{Filamentation of a Relativistic Proton Bunch in Plasma}, Phys. Rev. E (2024)~\cite{liviofilamentation}
    \item \textit{Hosing of a long relativistic particle bunch in plasma}, \textbf{Phys. Rev. Lett.} (2024)~\cite{tatiana}
    \item \textit{Development of the Self-Modulation Instability of a Relativistic Proton Bunch in Plasma}, Phys. Plasmas (2023)~\cite{liviodev}
    \item \textit{Controlled Growth of the Self-Modulation of a Relativistic Proton Bunch in Plasma}, \textbf{Phys. Rev. Lett.} (2022)~\cite{livio} 
    \item \textit{Simulation and Experimental Study of Proton Bunch Self-Modulation in Plasma with Linear Density Gradients}, Phys. Rev. Accel. Beams (2021)~\cite{pablo}
    \item \textit{Transition between Instability and Seeded Self-Modulation of a Relativistic Particle Bunch in Plasma}, \textbf{Phys. Rev. Lett.} (2021)~\cite{fabian}
    \item \textit{Proton Bunch Self-Modulation in Plasma with Density Gradient}, \textbf{Phys. Rev. Lett.} (2020)~\cite{falk}
    \item \textit{Experimental Study of Wakefields Driven by a Self-Modulating Proton Bunch in Plasma}, Phys. Rev, Accel. Beams (2020)~\cite{marleneprab}
    \item \textit{Experimental Observation of Proton Bunch Modulation in a Plasma at Varying Plasma Densities}, \textbf{Phys. Rev. Lett.} (2019)~\cite{karl}
    \item \textit{Experimental Observation of Plasma Wakefield Growth Driven by the Seeded Self-Modulation of a Proton Bunch}, \textbf{Phys. Rev. Lett.} (2019)~\cite{marlene}
    \item \textit{Acceleration of electrons in the plasma wakefield of a proton bunch}, \textbf{Nature} (2018)~\cite{nature}
\end{itemize}

\subsection{Numerical Simulations}

In AWAKE several advanced numerical simulation tools are employed to model and understand the complex beam-plasma interactions in the experiment. %
These tools aid to design, predict, and analyze the experiments. %
In general, we find excellent agreement between experimental and numerical simulation results~\cite{gorn}. %

Early simulations explored the self-modulation~\cite{kumar} and the wakefield phase and amplitude~\cite{marianaphase,pukhov} control, supporting the design of the experiment. %
Simulations also allow to better understand and interpret the experimental results such as the effects of ion motion on the self-modulation instability~\cite{marleneionmotion}, the filamentation processes in mismatched drive beams~\cite{erwin}, etc.
As the experiment advances, simulations are important to inform the complex experimental design for example showing that a correctly matched witness bunch can be accelerated to high energies, with a narrow relative energy spread, and without significant emittance growth~\cite{veronica}. %

Looking towards first applications, simulations have proposed alternative injection strategies~\cite{earli,kudia}, developed methods to mitigate instabilities~\cite{mariana}, identified regimes which allow acceleration beyond the dephasing limit~\cite{lotovppfc}, and suggested approaches to enhance the luminosity of the AWAKE scheme~\cite{Alumi}. %

\section{Steps towards an Accelerator for Particle Physics Applications}
The main goal of all future AWAKE experiments is acceleration with control of the quality of the accelerated bunch. %
This will be achieved first, by separating the self-modulator and the accelerator plasmas, second, by injecting the electron bunch on the axis of the wakefields, and third, by scaling the length of the plasma source to reach larger values of the energy gain. %
    \begin{figure}[phtb]
    \centering
    \includegraphics[scale=0.5]{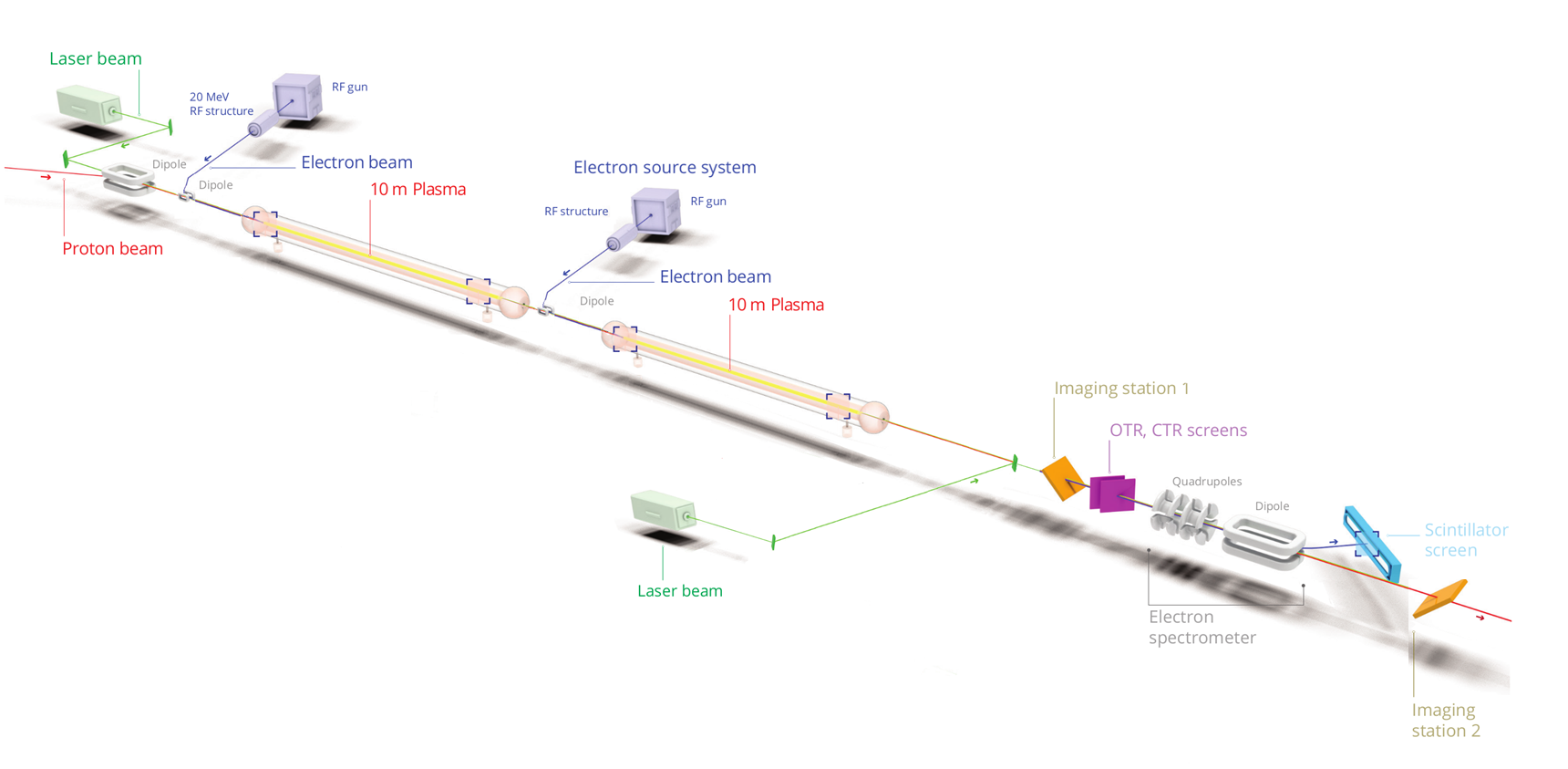}
    \caption{Layout of the experimental setup for the post-LS3 AWAKE experiments (2029-2033) showing in particular the two vapor/plasma sources and the new electron injector for injection of the accelerated bunch. 
    } 
    \label{fig:layout} 
    \end{figure}


\subsection{Optimizing Energy Gain and Bunch Quality}

Starting after CERN's Long Shutdown 3 (LS3), in a first phase of approximately two years, energy gain and bunch quality (charge, emittance, energy and relative energy spread) will be optimized both with RIF and electron bunch seeding in the self-modulator. %
These experiments will be performed with a 10\,m-long accelerator plasma in a laser-ionized rubidium vapor source following the self-modulator plasma (Fig.~\ref{fig:layout}). %

In plasma-based accelerators to control emittance and maintain a narrow energy spread during acceleration, the bunch must reach blowout of plasma electrons, must be matched to the focusing force of the pure column it therefore generates, and must properly load the wakefields. %
Numerical simulation results show that that for a plasma electron density of  $7\times$10$^{14}$\,cm$^{-3}$~\cite{veronica}, 
a bunch normalized emittance of 2\,mm-mrad, and an energy of 150\,MeV, the matched rms transverse size is 5.7\,\textmu m. %
Blowout and beam loading are reached with a charge of 100\,pC in a bunch with rms length of 60\,\textmu m. %

To produce this bunch, an innovative S-, X-band linac has been developed~\cite{linac}, %
that will be followed by a transfer line to deliver the electron bunch at the plasma entrance~\cite{rebecca}. %

Numerical simulation results show that evolution of the self-modulated proton bunch in the gap between the two plasmas leads to a decrease in the amplitude of the wakefields driven in the second plasma. %
The effect increase with gap length, and this length is minimized in the design of the injection region. %

\subsection{Scalability of Acceleration} 

In the second phase, the following two years ($\sim$2032-2033), the second plasma source will be replaced by a scalable discharge or helicon source, probably 20\,m long, to demonstrate the scalability in length of the source, and of the energy gain, still with an accelerated bunch of good quality. %
Much larger energy gains (50 to 200\,GeV) should then be possible by ''simply'' increasing the length of the accelerator source. %

\begin{figure}[phtb]
    \centering
    \includegraphics[scale=0.5]{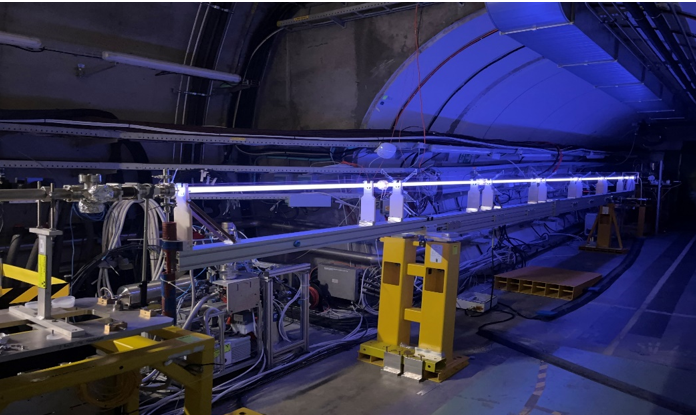}
    \caption{Ten meter-long discharge plasma source (DPS) installed to demonstrate that such a source can be used in an accelerator environment. %
    } 
    \label{fig:DPS2} 
    \end{figure}

Currently, helicon plasma sources (HPSs)~\cite{HPS,Stollberg_PPCF2024,Zepp_PoP2024,Granetzny_PoP2023} and discharge plasma sources (DPSs)~\cite{nuno,carolina} are under development in dedicated laboratories at CERN and collaborating institutes. %
By stacking units of RF antennas and magnetic field coils (HPS), or multiple sources with one cathode at high-voltage in the middle and a grounded electrode at each end (DPS), scalability of the length of the source can be obtained. %
A 10\,m-long DPS prototype (see Fig.~\ref{fig:DPS2}) was successfully tested in 2023 in the AWAKE facility, demonstrating the flexibility of the source for various experiments, including self-modulation studies~\cite{carolina}, ion-motion~\cite{marleneionmotion}, filamentation~\cite{liviofilamentation}, plasma light measurements, etc. %
These experiments have demonstrated that such a plasma source can be very successfully used in an accelerator environment. %

The current activities focus on the development of plasma diagnostics and their comparison, on discharge stability and reproducibility assessments for the DPS, and axial density profile and basic plasma studies for the HPS. %
The plasma source technology used for 
the scaling experiments will be decided in 2028. %

Plans for applications of the AWAKE acceleration scheme are detailed in separate documents \cite{PBC, alive, symmetry}. %

\section{Synergies}
The AWAKE experiment has fulfilled all planned milestones and experimental results so far, and has prepared a very clear plan from proof-of-concept to a technology that can provide beams for particle physics experiments. %
Moreover, the development of AWAKE technologies will have a profound impact on lepton facilities and colliders, based on both plasma wakefield acceleration, and on conventional technologies: 
\begin{itemize}
    \item The R\&D on plasma sources is common to other plasma-based colliders. %
    AWAKE will set the world-wide standard of very long plasma sources for electron acceleration in terms of density uniformity, reproducibility, tunability, and stability. %
    Other plasma accelerator concepts such as HALHF~\cite{halhf}, ALiVE~\cite{alive} and  other multi-TeV collider~\cite{lcv,alegro,10tev}, will benefit from the AWAKE plasma source developments. %
    \item Controlled external injection of the electron bunch into the plasma, and large energy gain whilst controlling bunch emittance, are two key challenges for plasma accelerators and are of significant interest to the entire field. %
    Several international facilities such as EuPRAXIA, FLASHForward, and FACET-II address single plasma stage issues. %
    AWAKE complements these efforts taking an integrated approach to developing a plasma-based accelerator for particle physics applications. %
    \item Efforts with any future CERN project involving electron colliders (e.g., FCC-ee, CLIC, LCvision) and facilities (FCC-ee injector test-stand, etc.) will be ramped up as AWAKE trains experts on electron acceleration, addressing a variety of beam dynamics and technological challenges, as demonstrated by a large number of high-level publications, PhD theses defended and prestigious prizes awarded. %
    \item The first experiments proposed to follow Run 2 are likely to be the first to benefit the HEP community, and hence will be an important opportunity to connect the advanced accelerator community with the traditional, high energy physics community. %
\end{itemize}

\section{Conclusions}

As demonstrated in this document, AWAKE has made remarkable progress towards establishing the concept of a plasma wakefield accelerator driven by a high-energy, self-modulated proton bunch. %
There is a clear plan to demonstrate external injection of an electron bunch in an accelerator plasma to produce a multi-GeV energy bunch with parameters suitable for applications. 
The plan also includes demonstration of the scalability of the acceleration process to produce multi-tens of GeV electron bunches before CERN's Long Shutdown 4 (LS4) starting in 2034.
At that time, AWAKE will be in a position to propose particle physics experiments that can be performed with these bunches, e.g., strong field QED, and fixed-target experiments. %
AWAKE experiments before LS4 can therefore be seen as a demonstrator for a plasma-based accelerator for application to particle and high-energy physics. %

In addition to this programmatic goal, the topics addressed by AWAKE are extremely synergetic with other projects, and with other plasma-based accelerators, thereby advancing the field in general. %

Therefore, considering the remarkable progress and the clear plans for further development presented in this document, we are of the opinion that AWAKE, as well as similar advanced accelerator R\&D be strongly supported by the European Strategy for Particle Physics Update. 


\



\begin{thebibliography}{99}


\bibitem{chen}P.~Chen, et al., Phys. Rev. Lett. 54, 693 (1985). 

\bibitem{rosenzweig}J.~B.~Rosenzweig et al., Phys. Rev. Lett. 61, 98 (1988). 

\bibitem{blumenfeld}I.~Blumenfeld et al., Nature 445, 741-744 (15 February 2007).

\bibitem{litos}M.~Litos et al., Nature 515, 92 (2014).

\bibitem{lindstrom}C.~Lindstrøm et al., Nat Commun 15, 6097 (2024).

\bibitem{kumar}
K.~V.~Lotov, Proc. 6th European Particle Accelerator Conference, 806-808 (1998), N.~Kumar et al., Phys. Rev. Lett. \textbf{104}, 255003 (2010). 

\bibitem{lotovEwb} K.~Lotov, Phys. Rev. Lett. 112, 194801 (2014).

\bibitem{lotovstep} 
A.~Caldwell and K.~V.~Lotov, 
Phys. Plasmas \textbf{18}, 103101 (2011).

\bibitem{veronica}V.~K.~Berglyd~Olsen et al., Phys. Rev. Accel. Beams 21, 011301 (2018).

\bibitem{PBC} 
R.~Alemany-Fernandez et al., CERN-PBC-Report-2025-03, submitted to 2026 European Particle Physics Strategy Update.  

\bibitem{symmetry} 
E.~Gschwendtner et al. (AWAKE Collab.), Symmetry, 14(8), 1680 (2022).

\bibitem{alive}
A.~Caldwell et al., arXiv:2503.21669, submitted to 2026 European Particle Physics Strategy Update.

\bibitem{karl}
AWAKE~Collaboration, Phys. Rev, Lett. \textbf{122}, 054802 (2019).

\bibitem{marlene}
M.~Turner et al. (AWAKE Coll.), Phys. Rev, Lett. \textbf{122}, 054801 (2019).

\bibitem{lotovnoise} K. V. Lotov et al., Phys. Rev. ST Accel. Beams 16, 041301 (2013). 

\bibitem{fabian}
F.~Batsch et al. (AWAKE Coll.), Phys. Rev. Lett. \textbf{126}, 164802 (2021).

\bibitem{livio}
L.~Verra et al. (AWAKE Coll.), Phys. Rev. Lett., 129, 024802 (2022).

\bibitem{nature}
AWAKE~Collaboration, Nature \textbf{561}, 363 (2018).

\bibitem{royal} 
E.~Gschwendtner et al. (AWAKE Coll.), Phil. Trans. R. Soc. A. \textbf{377}, 20180418 (2019).

\bibitem{pablothesis}
P.~I.~Morales Guzmán, PhD Thesis, TUM (2023).

\bibitem{pukhov}
A.~Pukhov et al., Phys. Rev. Lett. \textbf{107}, 145003 (2011).

\bibitem{schroeder}C.~B.~Schroeder et al., Phys. Rev. Lett. \textbf{107}, 145002 (2011). 

\bibitem{falk}
F.~Braunmueller et al. (AWAKE Coll.), Phys. Rev. Lett. \textbf{125}, 264801 (2020).

\bibitem{pablo}
P.~I.~Morales Guzmán et al. (AWAKE Coll.), Phys. Rev. Accel. Beams \textbf{24}, 101301 (2021).

\bibitem{marleneionmotion}
M.~Turner et al. (AWAKE Coll.), accepted in Phys. Rev. Lett. (Feb. 2025), arXiv:2406.16361 [physics.plasm-ph].

\bibitem{liviofilamentation}
 L.~Verra et al., (AWAKE Collab.), Phys. Rev. E 109, 055203 (2024). 

\bibitem{tatiana}
T.~Nechaeva et al. (AWAKE Coll.), Phys. Rev. Lett. \textbf{132}, 075001 (2024).

\bibitem{liviodev}
L.~Verra et al. (AWAKE Coll.), Phys. Plasmas 30, 083104 (2023).

\bibitem{marleneprab}
M.~Turner et al. (AWAKE Coll.), Phys. Rev, Accel. Beams \textbf{23}, 081302 (2020).

 \bibitem{gorn}
 A.~A.~Gorn et al. (AWAKE Collaboration), Plasma Phys. Control. Fusion 62, 125023 (2020).

 \bibitem{marianaphase}
 M.~Moreira, Phys. Rev. Accel. Beams \textbf{22}, 031301 (2019).

\bibitem{erwin} 
E. Walter et al., Phys. Rev. E \textbf{110}, 035208 (2024).

 \bibitem{earli}
 S.~Marini et al., Phys. Rev. Accel. Beams \textbf{27}, 063401 (2024).

\bibitem{kudia}
V.~Khudiakov and A. Pukhov, Phys. Rev. E \textbf{105}, 035201 (2022).

 \bibitem{mariana}
 J.~Vieira et al., Phys. Rev. Lett. 112, 205001 (2014), M.~Moreira et al., Phys. Rev. Lett. \textbf{130}, 115001 (2023).

 \bibitem{lotovppfc}
 K.~V.~Lotov and P.~V.~Tuev, Plasma Phys. Control. Fusion 63, 125027 (2021).

 \bibitem{Alumi}
 J.~P.~Farmer and G.~Zevi Della Porta, Phys. Rev. Res. 7, L012055 (2025). 

\bibitem{linac}M.~D.~Kelisani et al., Nuclear Inst. and Methods in Physics Research, A 982 164564, (2020).

\bibitem{rebecca}
R.~Ramjiawan et al., NIMA 1049, (2023), 168094.

\bibitem{HPS}
B.~Buttenschön et al., Plasma Phys. Control. Fusion 60, 075005 (2018)

\bibitem{Stollberg_PPCF2024}
C.~Stollberg et al., Plasma Phys. Control. Fusion \textbf{66}, 115011 (2024).

\bibitem{Zepp_PoP2024}
M.~Zepp et al., Phys. Plasmas \textbf{31}, 070704 (2024).

\bibitem{Granetzny_PoP2023}
M.~Granetzny et al., Phys. Plasmas \textbf{30}, 120701 (2023).

\bibitem{nuno}
N.~Torrado et al., IEEE TRANSACTIONS ON PLASMA SCIENCE, VOL. 51, NO. 12, (2023).

\bibitem{carolina}
C.~Amoedo et al., AWAKE Collaboration, in preparation (2025).

\bibitem{halhf}
B.~Foster et al., arXiv:2503.19880v1, submitted to 2026 European Particle Physics Strategy Update.

\bibitem{10tev}
S.~Gessner et al., arXiv:2503.20214v1, submitted to 2026 European Particle Physics Strategy Update.

\bibitem{lcv}
H.~Abramovicz et al., arXiv:2503.19983v1, submitted to 2026 European Particle Physics Strategy Update.

\bibitem{alegro}
B.~Cros, P.~Muggli et al.,  arXiv:1901.08436v2, submitted to 2026 European Particle Physics Strategy Update.


\end{thebibliography}
\end{document}